\documentclass{webofc}
\usepackage[varg]{txfonts}   

\newcommand{\tth}     {{ {\scshape The Three Hundred}}}
\newcommand{\sage}    {{ {\sc Sage}}}
\newcommand{\tkdmo}   {{ \textbf{{\scshape3k-dmo}}}}
\newcommand{\skdmo}   {{ \textbf{{\scshape7k-dmo}}}}
\newcommand{\tksage}  {{ \textbf{{\scshape3k-Sage}}}}
\newcommand{\sksage}  {{ \textbf{{\scshape7k-Sage}}}}

\newcommand{\fkdmo}   {{ \textbf{{\scshape15k-dmo}}}}
\newcommand{\tkgizmo} {{ \textbf{{\scshape3k-gizmo}}}}
\newcommand{\skgizmo} {{ \textbf{{\scshape7k-gizmo}}}}

\newcommand{\modot}{$\mathrm{M}_\odot$}
\newcommand{\h}{$h^{-1}$}

\begin{document}

\title{Galaxy catalogs from the {\sc Sage} Semi-Analytic Model calibrated on {\sc The  Three Hundred} hydrodynamical simulations: A method to push the limits toward lower mass galaxies in dark matter only clusters simulations}
%
%

\author{
        \lastname{Jonathan S. G\'omez}\inst{1}\fnsep\thanks{e-mail: j.s.gomez.u@gmail.com}
        \and
        \firstname{G.} \lastname{Yepes}\inst{1}
        \and
        \firstname{A.} \lastname{Jim\'enez Mu\~noz}\inst{1,2}
        \and
        \firstname{W.} \lastname{Cui}\inst{1,3}
        }

\institute{
        Departamento de F\'isica Te\'orica and CIAFF, M\'odulo 8, Facultad de Ciencias, Universidad Aut\'onoma de Madrid, 28049 Madrid, Spain 
        \and
        Univ. Grenoble Alpes, CNRS, Grenoble INP, LPSC-IN2P3, 53, avenue des Martyrs, 38000 Grenoble, France
        \and
        Institute for Astronomy, University of Edinburgh, Edinburgh, UK
        }

\abstract{
The new generation of upcoming deep photometric and spectroscopic surveys will  allow us to measure the astrophysical properties of faint galaxies in massive clusters. This would demand to produce  simulations of galaxy clusters with better mass resolution than the ones available today if we want to  make comparisons between the upcoming observations and predictions of cosmological models.  But producing full-physics hydrodynamical simulations of  the most massive clusters is not an easy task. This would involve billions of computational elements to reliably resolve low mass galaxies  similar to those measured in observations.  On the other hand,  dark matter only simulations of cluster size halos can be done with much larger mass resolution but at the cost of  having to apply a model that populate galaxies within  each of the subhalos  in these simulations. In this paper we present the results of a new set of  dark matter only simulations with different mass resolutions  within the \tth~project. We  have  generated catalogs of galaxies with stellar and luminosity   properties  by applying  the \sage~ Semi-Analytical Model of galaxy formation.  To obtain the catalogs consistent with  the results from  hydrodynamical simulations, the internal physical parameters of \sage~ were calibrated with the Particle Swarm Optimization  method  using  a subset of  full-physics runs   with the same mass resolution than the dark matter only ones.
}

\maketitle

\section{Introduction}
\label{intro}
The new generation of upcoming deep surveys like {\sc Euclid} \cite{euclidredbook}, {\sc 4most} \cite{RSdeJong} and {\sc Weave} \cite{jinweave},  will detect galaxies within clusters down to very faint magnitudes. This will help to understand the physics of galaxy formation and evolution in these high density environments and will  also  allow to derive more accurate mass estimates of the total mass of clusters from their galaxy content, which is key to  be able to put constrains on  the cosmological parameters of the Universe. But, at the same time, it will also demand new methods to perform  cosmological simulations  of cluster size objects with better mass resolutions  to be able to contrast the  predictions from the theoretical models  with those observed by the new surveys. The full-physics hydrodynamical simulations of massive clusters generated by the ‘zoom-in’ technique adopted in \tth\footnote{https://www.the300-project.org.} \cite{cui2018three} project offer the perfect laboratory for a comparison with current  surveys but  they have a mass resolution not enough to  resolve dark matter sub haloes with $10^{11}$ \h \modot~ and below. Their hosted galaxies have magnitudes  that are fainter than those coming from upcoming (e.g. Euclid will reach magnitudes $m_{\mathrm{H}} \sim 24$) surveys.  For this reason, a new  generation of hydrodynamical simulations with higher mass resolution of \tth~ clusters is being carried out but given the high computational cost, only 5 zoomed regions, out of a total of 324 regions,  have been completed so far.  On the other hand,  dark matter only versions of the full data set at high resolution   have already been completed using much less computational resources. Therefore, we can make an efficient use of these simulations if we can emulate the observational properties of the galaxies within each of the dark subhalos that mimic those coming from the full-physics hydrodynamical simulations  at our disposal. To do this, we use as an emulator  the publicly available \sage~ Semi-Analytic Model (SAM) \cite{croton2006} of galaxy formation, and calibrate their free parameters against observables from the hydrodynamical simulations.

\begin{table}[h]
\scriptsize
\centering
\caption{\tth~simulations.}
\label{table1}
\begin{tabular}{ccccc}
\hline
\hline
Name     & N particles & DM particle mass [\h \modot] & DM halo resolution (100p$^a$) [\h \modot] & N regions \\ \hline
\hline
3K-DMO   & $3840^3$    & $1.5 \times 10^9$              & $10^{11}$                                   & 324       \\
3K-GIZMO & $3840^3$    & $1.5 \times 10^9$              & $10^{11}$                                   & 324       \\ \hline
7K-DMO   & $7680^3$    & $1.8 \times 10^8$              & $10^{10}$                                   & 324       \\
7K-GIZMO & $7680^3$    & $1.8 \times 10^8$              & $10^{10}$                                   & 5         \\ \hline
15K-DMO  & $15360^3$   & $2.3 \times 10^7$              & $10^{9}$                                    & 3         \\ \hline
\multicolumn{5}{l}{$^a$ : 100 dark matter particles to form a halo.}
\end{tabular}
\end{table}

\section{Data}
\label{data}

\subsection{{\sc The Three Hundred} dataset}
Our data set is based on the 324 zoomed regions simulated within \tth~ project   with different physics flavours, including the newest runs with dark matter only.   They were created starting from the Dark Matter Only (DM-Only) MultiDark Simulation \citep{giocoli2016multidarklens} (MDPL2), which consist of a 1 \h Gpc cube containing $3840^3$ dark matter particles (DM) each with a mass of $1.5 \times 10^9$ \h \modot. The initial conditions were generated by  identifying the Lagrangian regions  of all the particles lying within a spherical region of 15 \h Mpc centred around each of the 324 most massive clusters   of MDPL2 at $z=0$.  Within this Lagrangian region,  high resolution dark matter and gas particles were populated  while dark matter particles of increasing mass levels  fill the  full box  outside the zoomed area, describing  the global gravitational field. \tth~Collaboration performed these simulations in five different variants based on the GIZMO-SIMBA \cite{cui2022} code, which can generate both DM-Only and hydrodynamical simulations. Depending mainly on the physics and resolutions, the five variants of \tth~ Simulations we are using in this work are: 1)\tkgizmo: Full-physics hydrodynamics zoomed simulations of  324 regions with a DM particle resolution of $1.5 \times 10^9$ \h \modot, 2)\tkdmo: DM-only simulations at the same resolution as the \tkgizmo~simulations and 324 regions are available, 3)\skgizmo: Physics as \tkgizmo~but at high resolution which contain $7680^3$ DM particles (twice particles per dimension than \tkdmo~or\tkgizmo) with a DM particle mass of $1.8 \times 10^8$ \h \modot. For these, we only have 5 available regions given its high computational cost, 4)\skdmo: DM-only simulations at high resolution as \skgizmo~ and 324 regions are available, 5)\fkdmo: DM-only simulations as \skdmo~ but at very high resolution which contain $15360^3$ DM particles (twice particles per dimension than \skdmo~or\skgizmo) with a DM particle mass of $2.3 \times 10^7$ \h \modot. For these simulations we only have 3 regions at the moment. For a global summary of the \tth~simulations see Table \ref{table1}.

\subsection{Semi-Analytic Model of Galaxy formation and Evolution: {\sc Sage}}
 In this work we use the \sage~ SAM  \cite{croton2006} which has 14  internal parameters to model the physical process of the galaxies through merger trees of DM haloes. All these parameter values can be tuned  to calibrate the \sage~ galaxy properties with respect to different (observed or simulated) constrains. Considering that we have the whole 324 regions in \skdmo~simulations, we  will apply the  \sage~ model on  them (\sksage)  for  calibrating the  parameters of the model in order to obtain  results as close as possible to the 5 \skgizmo~ available regions. Once the \sage~ model is calibrated we then apply it to the full \skdmo~ dataset to obtain the galaxy properties of all the halos and subhalos in these simulations.

\subsection{{\sc Sage} calibration with {\sc Particle Swarm Optimization}}
\label{subsection:calibration}
\begin{figure}[h]
\centering
\includegraphics[scale=0.19]{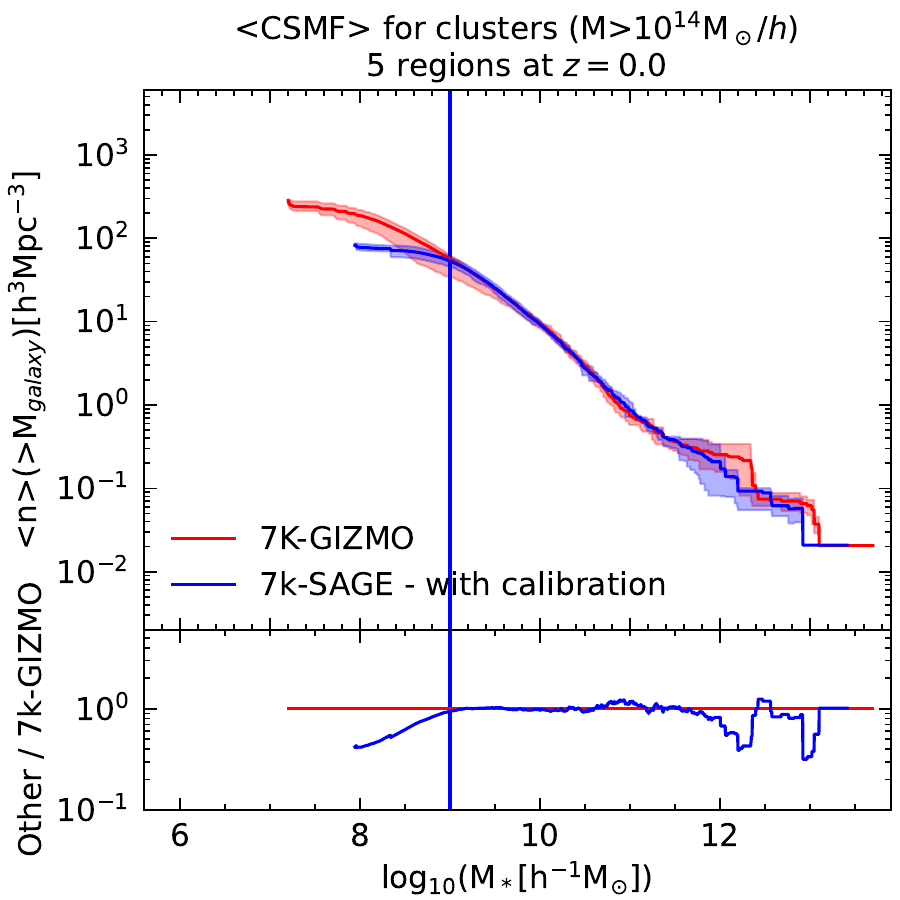}
\includegraphics[scale=0.19]{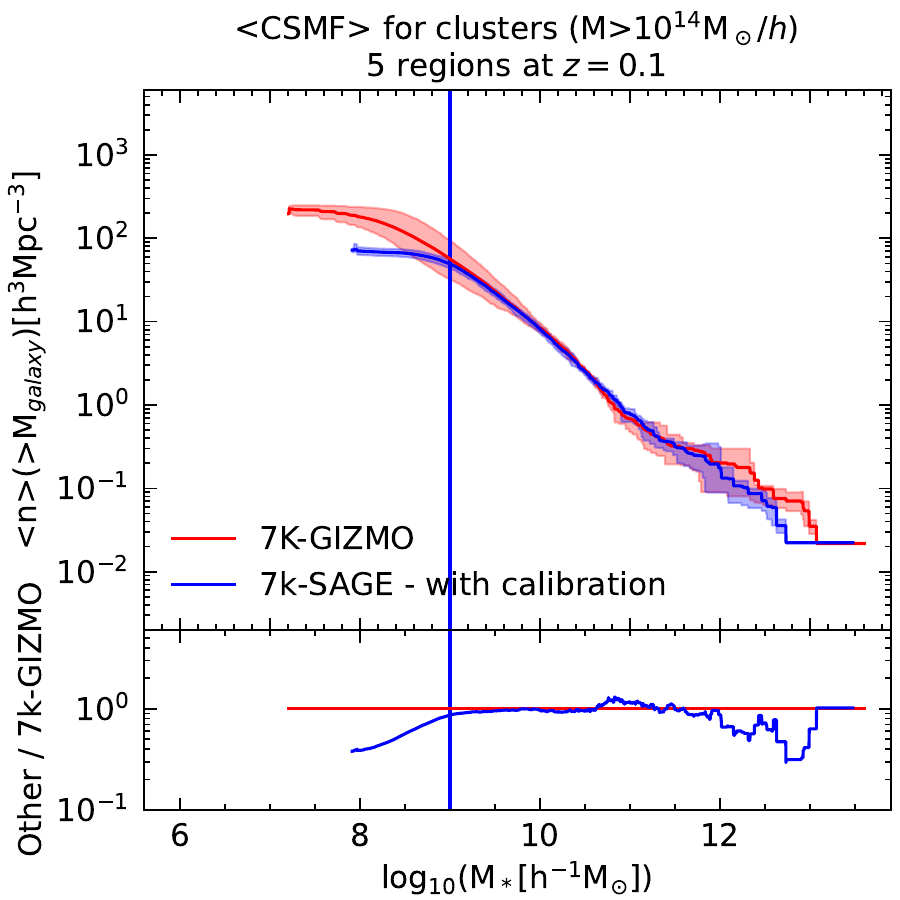}
\includegraphics[scale=0.19]{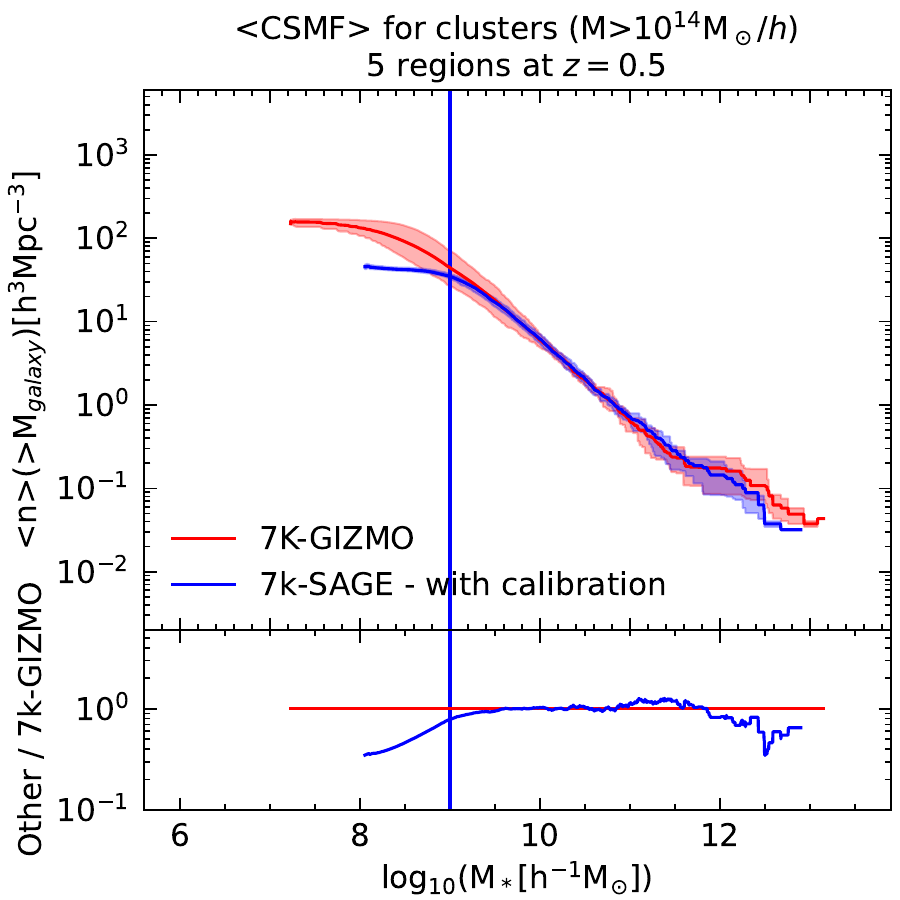}
\includegraphics[scale=0.19]{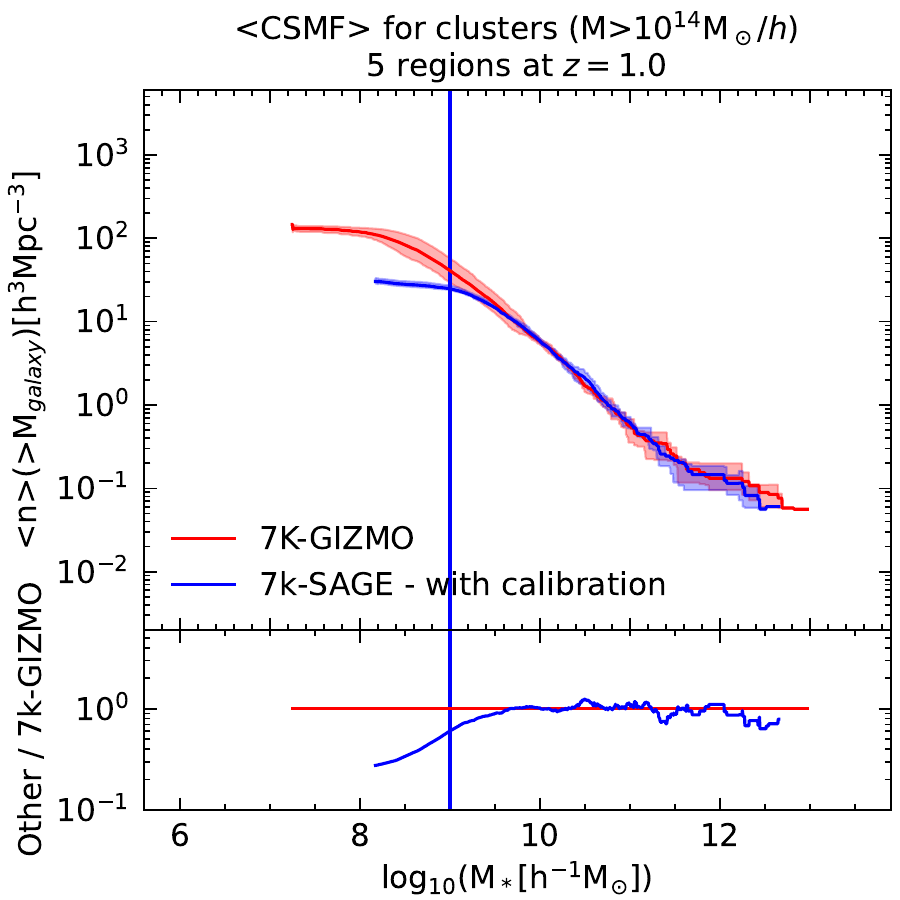}\\
\includegraphics[scale=0.19]{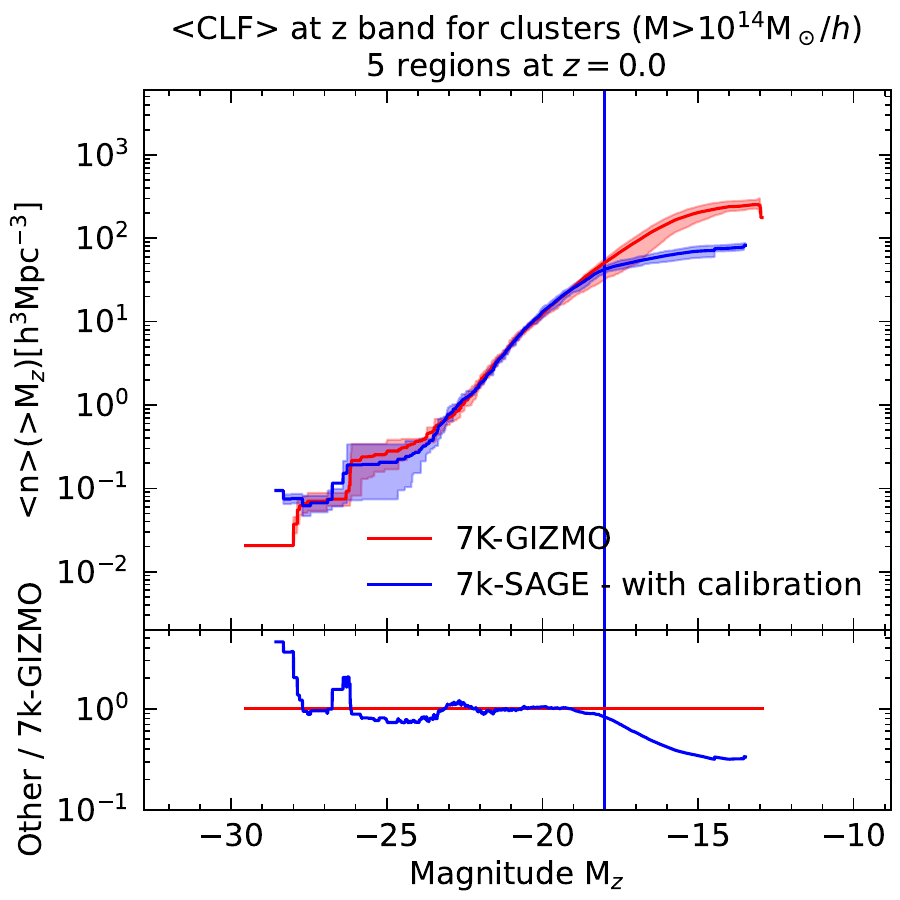}
\includegraphics[scale=0.19]{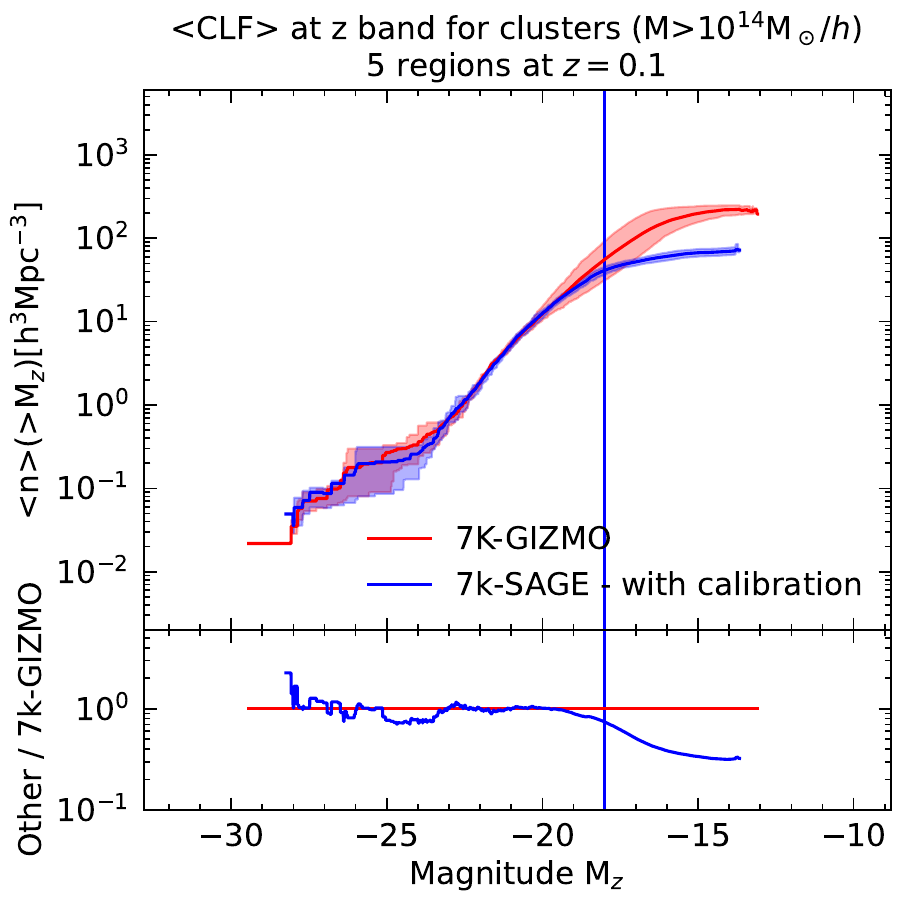}
\includegraphics[scale=0.19]{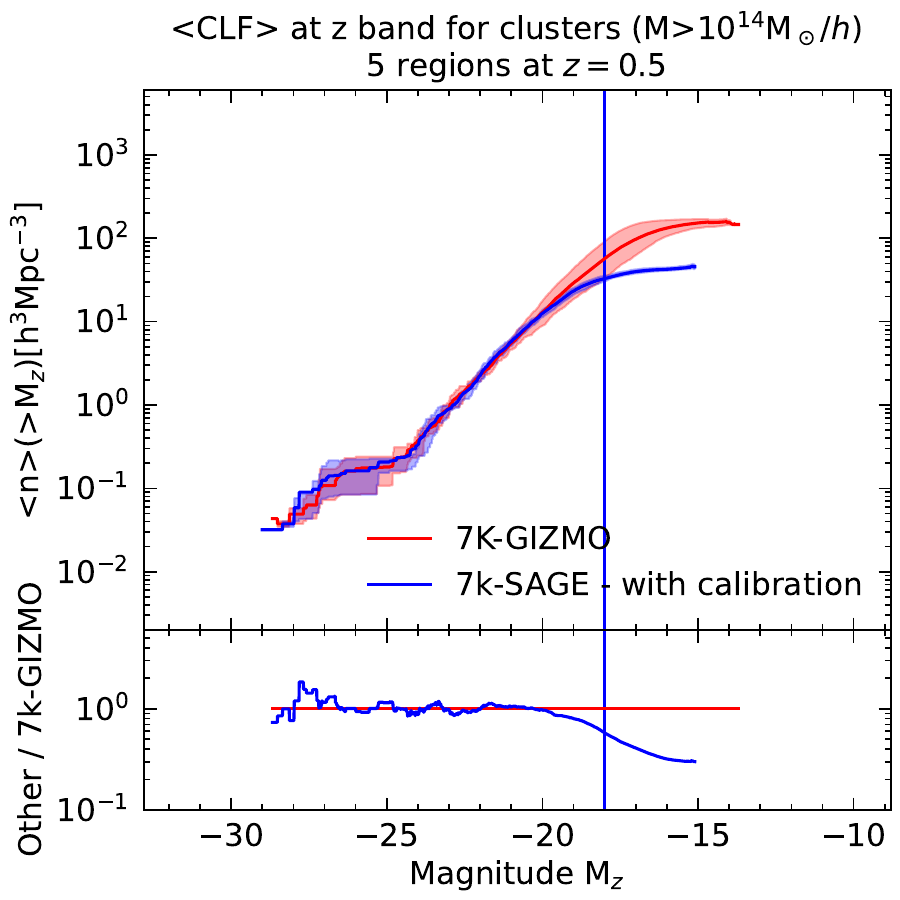}
\includegraphics[scale=0.19]{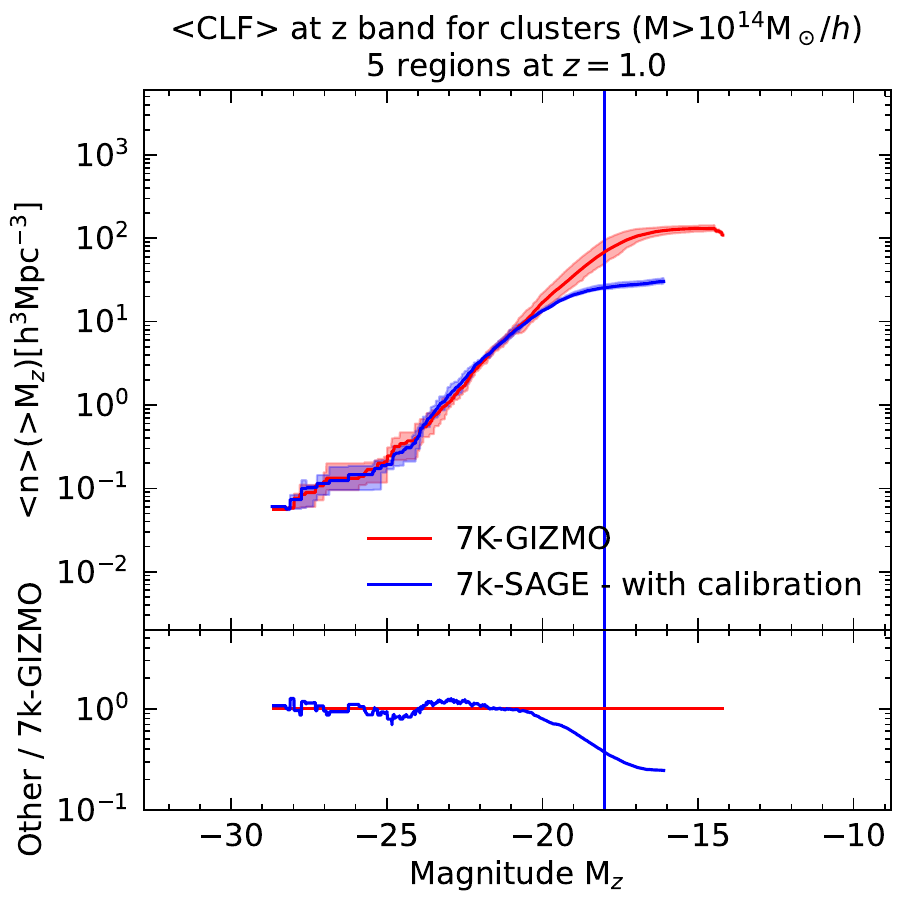}\\
\includegraphics[scale=0.19]{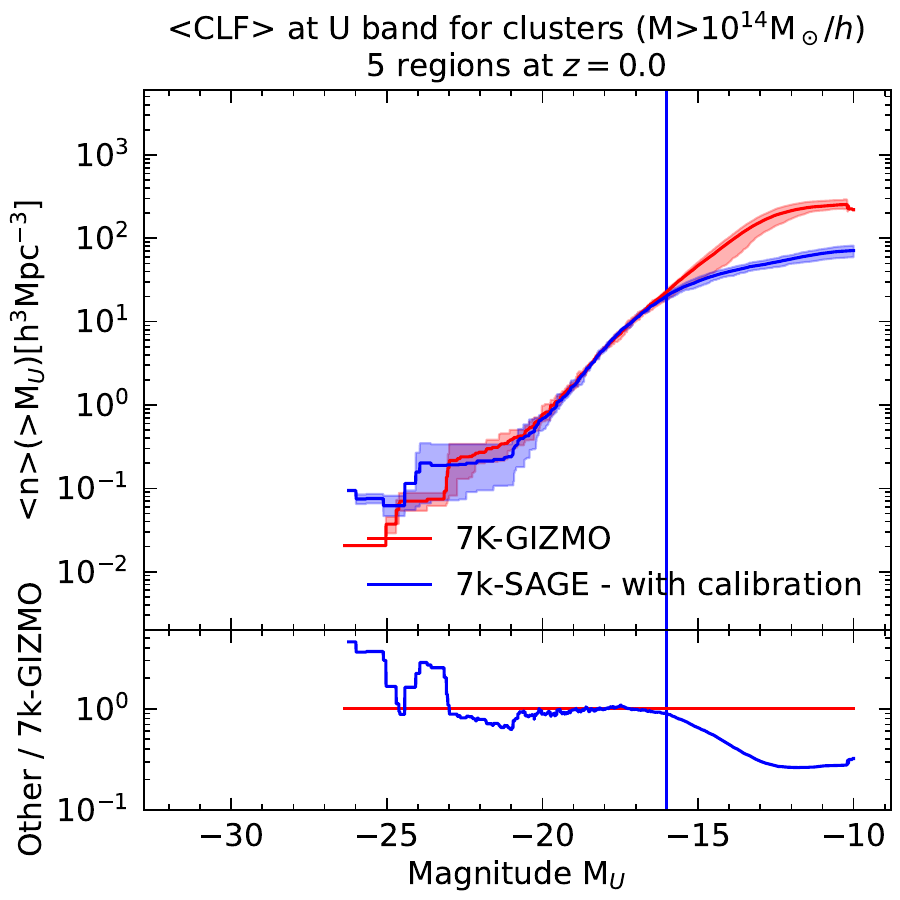}
\includegraphics[scale=0.19]{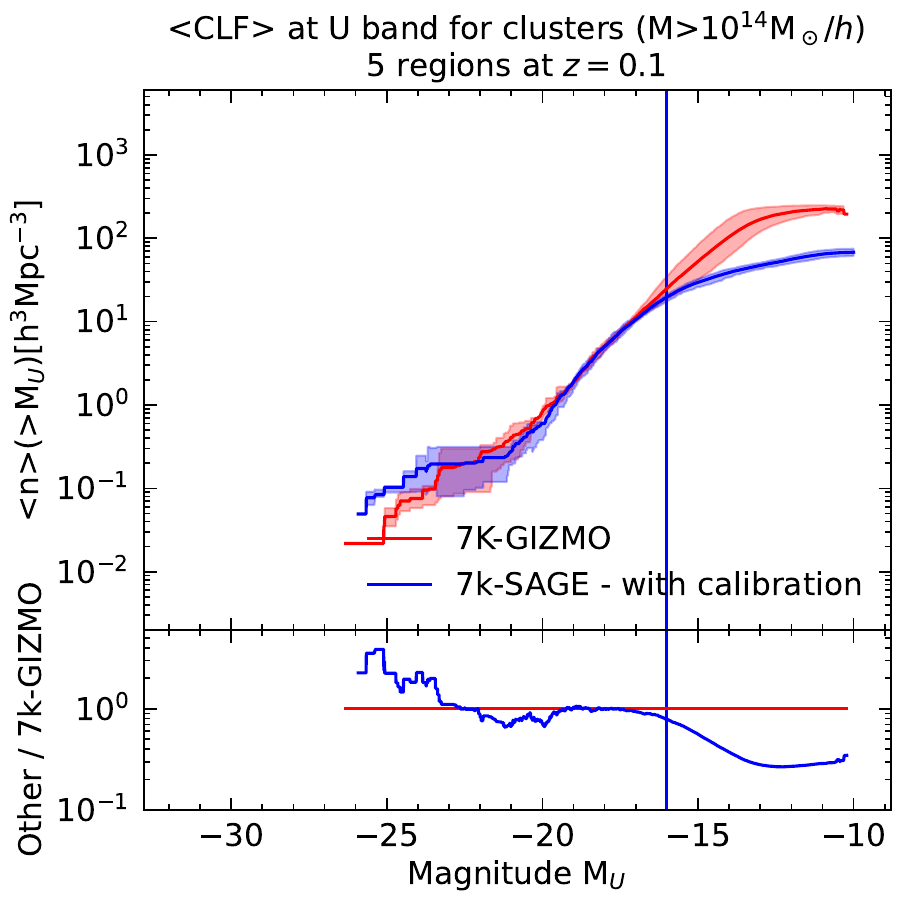}
\includegraphics[scale=0.19]{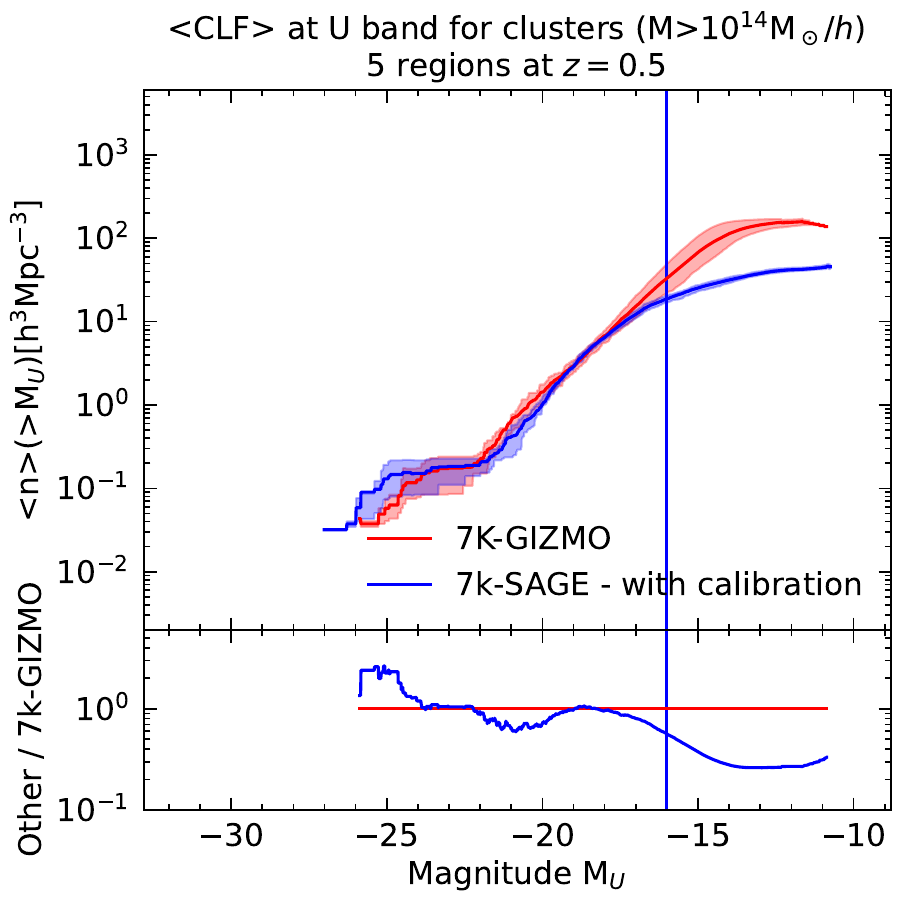}
\includegraphics[scale=0.19]{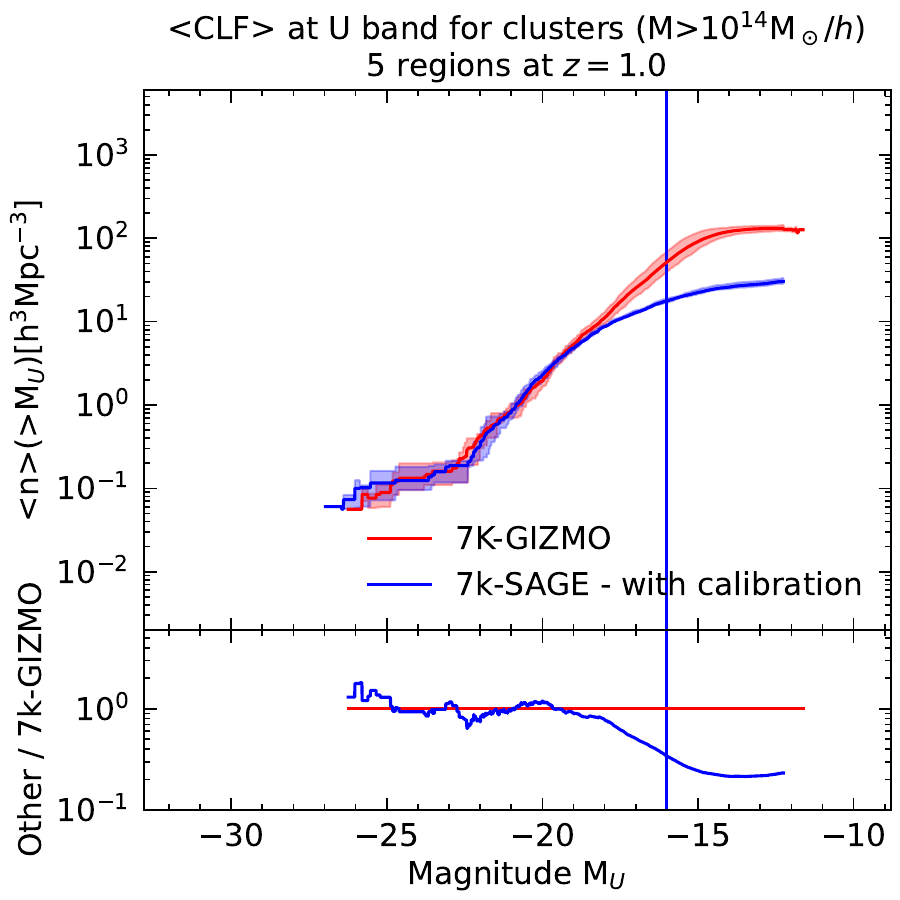}
\caption{Average cumulative stellar mass functions (<CSMF>; top panels), average cumulative luminosity functions for the z-band (<CL$_\mathrm{z}$F>; panels in the second row), and average cumulative luminosity functions for the U-band (<CL$_\mathrm{U}$F>; bottom panels) at 4 redshifts: $z=0$, $z=0.1$, $z=0.5$, and $z=1$ (left to right panels). These cumulative functions have been  used to calibrate \sksage~(blue line) with respect to \skgizmo~(red line) with PSO using all clusters with M$_{halo} > 10^{14}$ \h \modot~ in the 5 coincident regions of \skdmo~ and \skgizmo. The filled color regions correspond to the 1-$\sigma$ error and the vertical blue line represents  the limit used in the calibration procedure (minimum limit for stellar mass and maximum limit for luminosity). For each panel, the ratio of the average cumulative functions over that from \skgizmo is shown at the bottom.}
\label{fig:calibration}
\end{figure}

To calibrate \sage~ we use the {\sc Particle swarm optimization} (PSO) \cite{pso1995} method (see \cite{ruiz2015} for its first use to calibrate a SAM) which varies the internal parameters of \sksage~ efficiently and finds the optimal values that  minimize  the difference between   galaxy properties  between  \sksage~ and \skgizmo. PSO is a computational optimization technique in multi-dimensional spaces that is  based on the movement of particles represented by  their  positions  in the parameter space. In our case, each  PSO particle correspond to given values of the parameters of a  \sage~ run.  Using several of these particles, PSO works like  birds moving  in a swarm, hence the name. They  can profit from the common experience of all members of the swarm when looking for food. This search method is at least 30 times faster than the standard Monte Carlo minimization techniques (see \cite{mcmc2008} and \cite{mcmc2009}).

In this work, we  take as calibrators, the average cumulative stellar mass function (<CSMF>) and the z-band and U-band average cumulative luminosity function (<CLF>) of galaxies  within all  clusters more massive than  at  least $10^{14}$ \h \modot. We  compute these observable  from the 5 available regions of  \skgizmo -simulations and compute them also in   their corresponding   DM only version \skdmo ~simulations using \sage. The galaxy luminosities were calculated by the Stellar Population Synthesis model {\sc Stardust} \cite{stardust1999} for both \sksage~ and \skgizmo~. We calibrate these 3 galaxy properties at 4 different redshifts: $z=0$, $z=0.1$, $z=0.5$, and $z=1$ because we want to be able to reproduce the star formation history from hydrodynamical simulations. Finally, the \sksage~ calibration was done  considering simultaneously the 3 galaxy properties at 4 different redshifts (between 0 and 1) in order to have a consistent evolution of these  properties with respect to the hydrodynamical results. In Fig \ref{fig:calibration} we show the final \sksage~calibration considering a minimum stellar mass and maximum luminosity limit when we calibrate with PSO because there are not enough haloes in the \skdmo~ simulations compared to \skgizmo~ since in hydrodynamical  simulations of galaxy clusters there are more  low mass subhalos  because baryonic matter helps them to  survive \cite{dolag09}

\begin{figure}[h]
\centering
\includegraphics[scale=0.42]{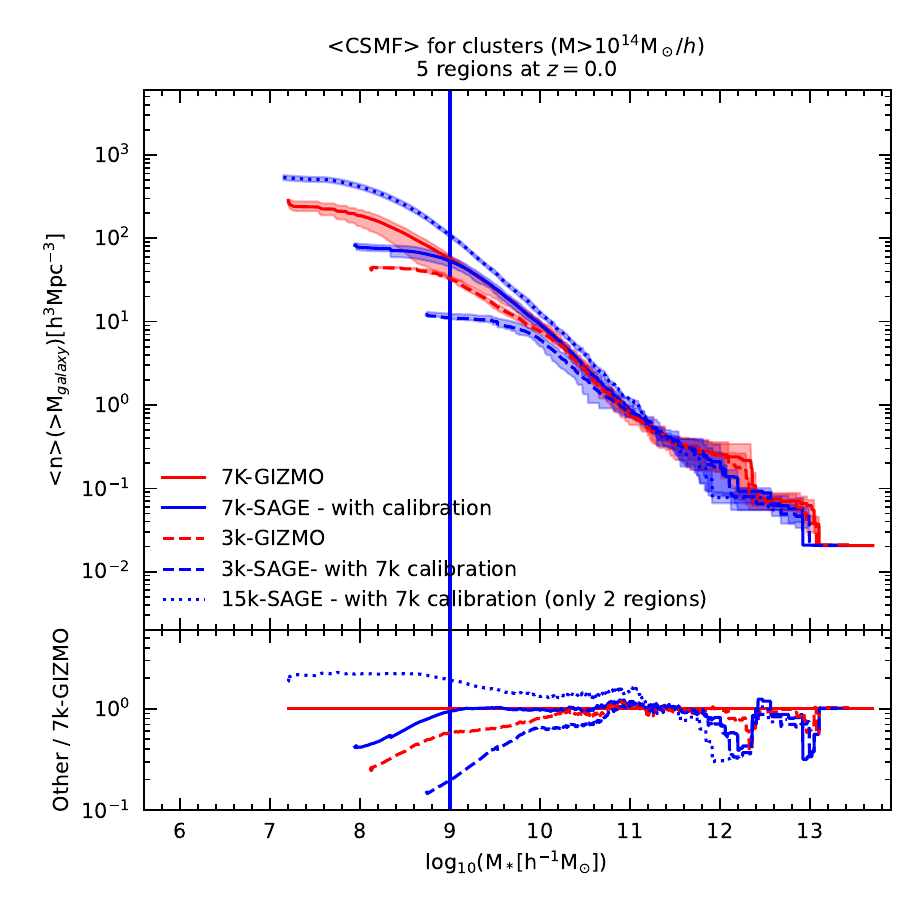}
\includegraphics[scale=0.42]{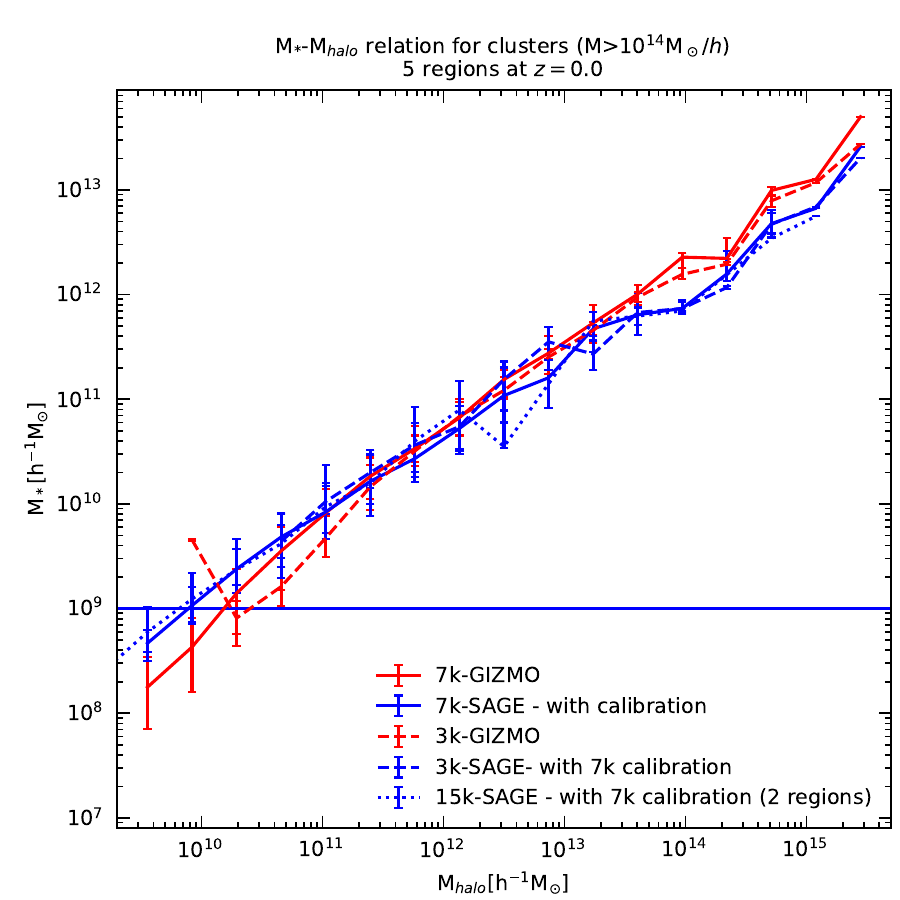}
\caption{Average cumulative stellar mass functions (<CSMF>, left panel) and stellar mass-halo mass relationship (right panel) at redshift $z=0$ for \tkgizmo~(dashed red line), \skgizmo~(solid red line), \sksage~calibrated (solid blue line), \tksage~(dashed blue line) with the parameters obtained in the \sksage~calibration and \fkdmo~(pointed blue line) using all clusters with M$_{halo} > 10^{14}$ \h \modot~ in the 5 common regions of \skdmo~ and \skgizmo. The filled colors in left panel correspond to the 1-$\sigma$ error and the vertical blue line to the minimum for stellar mass limit used in the calibration. The error bars in right panel correspond to the 1-$\sigma$ error.}
\label{fig:5regions-3resolutions}
\end{figure}

\begin{figure}[h]
\centering
\includegraphics[scale=0.42]{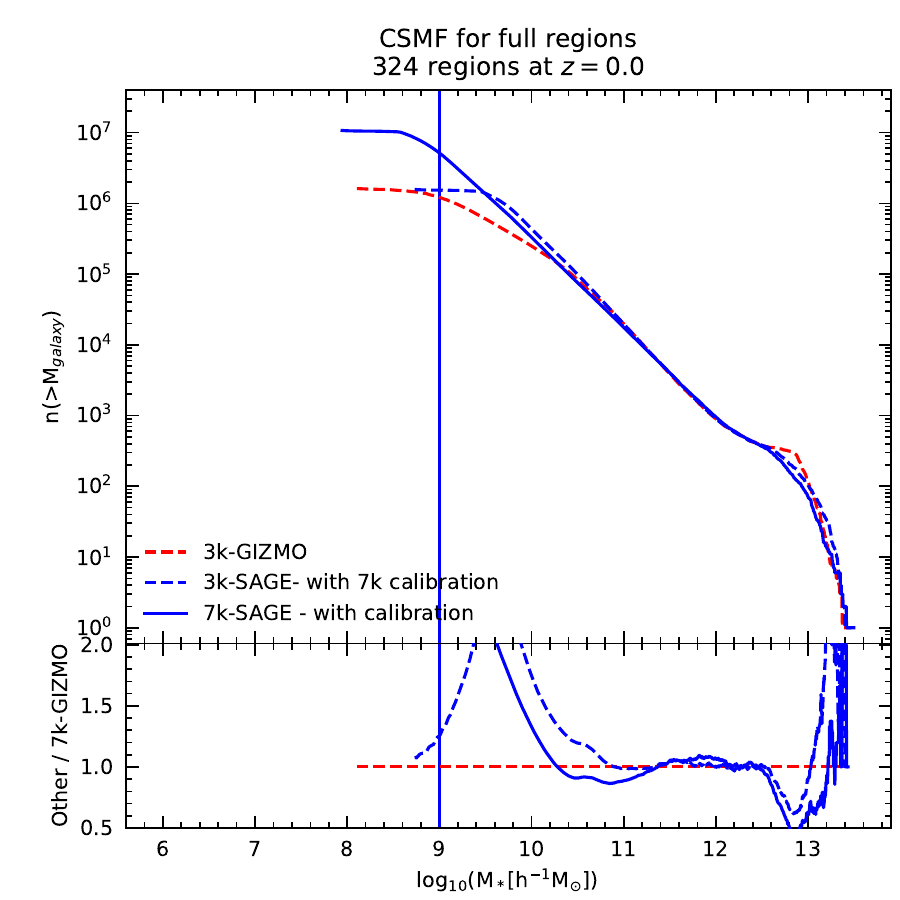}
\caption{cumulative stellar mass functions (CSMF) at redshift $z=0$ for \tkgizmo~(dashed red line), \sksage~calibrated (solid blue line), \tksage~(dashed blue line) with the parameters obtained in the \sksage~calibration using all available regions (see Table \ref{table1}). Vertical blue line corresponds to the minimum for stellar mass limit used in the calibration.}
\label{fig:324regions-2resolutions}
\end{figure}

\section{Results}
Considering the above mentioned calibration of \sage~parameters in Section \ref{subsection:calibration}, we can then apply the model on the other DM-Only simulation at different  resolutions: \tkdmo~ (low resolution) and \fkdmo~ (very high resolution). Thus we can  check whether  the galaxy properties obtained from the  calibrated \sage~are sensitive to the resolution of the dark matter halos. In the <CSMF> of  Fig. \ref{fig:5regions-3resolutions}-left we show all the versions of the simulations that we have considered (see Table \ref{table1} to see all data), where it seems that the application of the  \sksage~ calibration on different resolution DM-Only simulations, such as \tkdmo~ and \fkdmo, is  consistent before the minimum of the stellar mass limit. In addition, in the stellar mass - halo mass relationship of Fig. \ref{fig:5regions-3resolutions}-right we can see the consistency between the hydrodynamical simulations and the results of \sage~ calibrated on DM-Only simulation. Finally, with this calibration we can reach smaller galaxies by applying it in the DM-Only simulation at \fkdmo. As we can see in Fig. \ref{fig:324regions-2resolutions}, CSMF considering all the complete regions available in \tkgizmo, \tksage, \skgizmo~ and \sksage, this calibration opens the possibility to have new galaxy catalogs. We generated a catalog of 10 million galaxies from the \sage~ emulator from the 324 regions of \skdmo~ simulations, and 1 million galaxies through \sage~ from only 3 regions of \fkdmo~ simulations. These catalogs have a large number of galaxies because we have a large number of dark matter merger trees from all the host halos and their substructures.

\section{Conclusions}
\label{conclusions}

A galaxy emulator has been developed by \sage~ SAM applied to \tth~ DM-Only simulations to reproduce results from \skgizmo~ simulations such as stellar and luminosity properties of haloes at several redshifts (0 to 1). Given the low computational cost of DM-Only simulations + SAMs compared to hydrodynamical simulations we are able to push the limit towards lower mass galaxies in clusters simulations that can be very useful to make predictions for upcoming deep surveys like {\sc Euclid}, {\sc 4most} and {\sc Weave}.

In addition, the \sage~ calibration seems not to be sensitive to the resolution of the DM-Only simulation, so the calibration at \skdmo~ resolution can be safely applied in the \fkdmo~ higher resolution simulations. This emulator is amazingly flexible because we can add more galaxy properties and more redshifts to the calibration procedure with a  modest increase in computational time. Thus, this opens up the possibility of studying properties such as gas or supermassive black holes at $z > 5$, bringing us closer to the early universe of hydrodynamical simulations. Finally, a mock catalog of galaxies from the calibrated \sage~ is now ready to be used within \tth~ Collaboration.

\section*{Acknowledgements}
JSG acknowledges support from the Predoctoral contract `Formación de Personal Investigador' from the Universidad Autónoma de Madrid (FPI-UAM, 2021).

%
%

\end{document}